\documentstyle[aps,prl,floats,epsf]{revtex}
\tighten
\draft
\begin{document}

\twocolumn[\hsize\textwidth\columnwidth\hsize\csname
@twocolumnfalse\endcsname
\title{Neutrino magnetic moments, flavor mixing, and
the SuperKamiokande solar data}
\author{J.~F. Beacom and P. Vogel}
\address{Physics Department 161-33, California
Institute of Technology, Pasadena, California 91125, USA}
\date{July 15, 1999; revised September 28, 1999}
\maketitle

\begin{abstract}
We find that magnetic neutrino-electron scattering is unaffected by
oscillations for vacuum mixing of Dirac neutrinos with only diagonal
moments and for Majorana neutrinos with two flavors.  For MSW mixing,
these cases again obtain, though the effective moments can depend on
the neutrino energy.  Thus, e.g., the magnetic moments measured with
$\bar{\nu}_e$ from a reactor and $\nu_e$ from the Sun could be
different.  With minimal assumptions, we find a new limit on
$\mu_{\nu}$ using the 825-days SuperKamiokande solar neutrino data:
$|\mu_{\nu}| \le 1.5\times 10^{-10} \mu_B$ at 90\% CL,
comparable to the existing reactor limit.
\end{abstract}


\vspace{0.5cm}]
\narrowtext


In the minimally-extended Standard Model, neutrinos of mass $m_\nu$
have tiny loop-induced magnetic moments $\mu_\nu \simeq 3 \times
10^{-19} \mu_B \; (m_\nu/{\rm 1\ eV})$, where $\mu_B$ is the Bohr
magneton.  In various extensions of the Standard Model,
larger magnetic moments can occur without large neutrino masses.  In
the presence of flavor mixing, the fundamental magnetic moments are
associated with the mass eigenstates (since either a boost or a
magnetic moment can be used to reverse the helicity).  In the mass
eigenstate basis, Dirac neutrinos can have diagonal or off-diagonal
(transition) moments, while Majorana neutrinos can only have
transition moments~\cite{Kayser,Shrock}.

In the current experiments, the effects of neutrino magnetic moments
can be searched for only in the recoil electron spectrum from
neutrino-electron scattering~\cite{Kyuldjiev,Engel}.  Below we
consider the interplay between magnetic moments and flavor mixing for
this process.  We show how magnetic moments can be defined for
beams that are initially neutrino flavor eigenstates.  In some
important cases these moments do not oscillate, i.e., they do not
depend on distance from the source.  However, in the presence of MSW
mixing, these defined flavor moments can differ from the vacuum case
and can depend on the neutrino energy, though not on the
distance.  As an illustration, we derive a new limit on the magnetic
moment from the SuperKamiokande (SK) solar neutrino
data~\cite{Fukuda}.

There are two incoherent contributions to neutrino-electron
scattering: weak scattering, which preserves the
neutrino helicity, and magnetic scattering, which reverses it.  Thus
the differential cross section is given by
\begin{eqnarray}
\frac{d \sigma}{d T} & = &
\frac{2 G_F^2 m_e}{\pi} 
\left[g_L^2 + g_R^2 \left(1 - \frac{T}{E_{\nu}}\right)^2
-g_L g_R \frac{m_e T}{E_{\nu}^2}\right]  \nonumber \\
& + & \mu_{\nu}^2 \frac{\pi \alpha^2}{m_e^2} \frac{1 - T/E_{\nu}}{T}\,.
\label{eq:dsig}
\end{eqnarray}
In Eq.~(\ref{eq:dsig}), $g_L = \sin^2\theta_W + 1/2$ for $\nu_e$, $g_L
= \sin^2\theta_W - 1/2$ for $\nu_\mu$ and $\nu_\tau$, and $g_R =
\sin^2\theta_W$ for all flavors (for antineutrinos, exchange $g_L$ and
$g_R$).  The magnetic moment $\mu_\nu$ is expressed in units of
$\mu_B$.  Magnetic scattering, the second term in Eq.~(\ref{eq:dsig}),
grows rapidly with decreasing electron recoil kinetic energy $T$.

In principle, there can be weak-magnetic interference effects.  There
is a negligible effect due to the fact that a massive neutrino
is not a helicity eigenstate~\cite{Grimus}.  Also, if the neutrinos
have a transverse polarization, the electron azimuthal angle
distribution can be affected~\cite{azi}; we ignore this case, as the
effects are presently unobservable.


{\bf Vacuum Mixing:\ }
The effects of flavor mixing on the weak scattering are well-known.
Whatever the composition of the neutrino beam, the different flavors
are in principle distinguishable and hence their cross sections
combine incoherently, weighted by the probabilities for the neutrino 
to be of each given flavor.  

We want to explore how neutrino oscillations affect the magnetic
scattering.  The shape of the electron recoil spectrum in
magnetic scattering is universal (the same for all mass eigenstates).
The only quantity that depends on the beam composition is the
effective magnetic moment $\mu_{\nu}$.  Let us assume that we begin
with a beam of electron neutrinos.  Under the usual oscillation
hypothesis such a beam propagates over the distance $L$ from its
source in vacuum according to
\begin{equation}
| \nu_e (L) \rangle = \sum_k U_{ek} e^{-iE_k L} | \nu_k \rangle \,,
\end{equation}
where $U_{ek}$ is an element of the unitary mixing matrix and $k$
labels the mass eigenstates.

Similarly to above, whatever the composition of the neutrino beam, the
different mass eigenstates are in principle distinguishable in the
magnetic scattering, and hence their cross sections combine
incoherently, weighted by the squares of the amplitudes for
the neutrino to be of
each mass after the scattering.  Then the combined cross section for
magnetic scattering has the form of Eq.~(\ref{eq:dsig}) with
magnetic moment squared $\mu_\nu^2$ given by
\begin{eqnarray}
\mu_e^2 & = & \sum_j \left| \sum_k U_{ek} e^{-iE_k L} \mu_{jk} \right|^2
\nonumber \\
& = &
\sum_j \sum_{kk'} U_{ek} U_{ek'}^* \mu_{jk} \mu_{jk'}^*
e^{-2\pi i L/L_{kk'}}\,,
\label{eq:mue}
\end{eqnarray}
where the summations $j,k,k'$ are over the mass eigenstates, and the
subscript $e$ labels the initial flavor.  We have made the usual
relativistic expansion and have defined the oscillation length
$L_{kk'} = 4 \pi E_{\nu} /\Delta m_{kk'}^2 {\rm\ for} ~k \ne k'$
(there is no $L$ dependent phase for $k=k'$).  The quantities
$\mu_{jk}$ in Eq.~(\ref{eq:mue}) are the fundamental constants (in
units of $\mu_B$) that characterize the coupling of the neutrino mass
eigenstates to the electromagnetic field.  The
summation over $j$ is outside the square because of the incoherence of
the cross sections for different final masses.  The expression for
$\mu_e^2$ simplifies in some important cases.

Let us assume first that the neutrinos are Dirac particles (with $n$
flavors) with only diagonal magnetic moments ($\mu_{jk} = \mu_j
\delta_{jk}$); this is the scenario used by the Particle Data 
Group~\cite{RPP}.
Then
\begin{equation}
\mu_e^2 = \sum_j |U_{ej}|^2\; |\mu_j |^2\,,
\label{eq:Dir}
\end{equation}
and there is no dependence on the distance $L$ or neutrino energy
$E_{\nu}$.  In this case one can characterize the magnetic
scattering  by the {\it initial} flavor index instead of the
mass indices.  (Hence we disagree with Ref.~\cite{Gninenko}, in
which the magnetic
scattering depends on the {\it final} flavor index, i.e., it
oscillates).  Measurements of all magnetic moments and mixing
parameters would allow extraction of the ``fundamental'' moments
$\mu_j$.

Next consider the case of Majorana neutrinos, and assume that only
two mass eigenstates are relevant.  Then
\begin{equation}
\mu_e^2 = |\mu_{12}|^2 (|U_{e1}|^2 + |U_{e2}|^2) = |\mu_{12}|^2\,,
\label{eq:Maj}
\end{equation}
which is not only independent of the source distance and the neutrino
energy, but also of the mixing angle. 

Under what circumstances does one have to worry about a dependence on
distance and neutrino energy, and in particular, when can such terms
be dominant?  Clearly, at least one term of the type $\mu_{jk} \times
\mu_{jk'}$ with $k \ne k'$ must be nonvanishing and as large as
$\mu_{jk}^2$ or $\mu_{jk'}^2$.  In other words, in the $3 \times 3$
matrix $\mu_{jk}$ there should be at least two comparable entries on
the same line (and in the same column).  For the Dirac case this
implies that at least one nondiagonal magnetic moment is as large as
the diagonal ones.  For the Majorana case it implies that $two$
different nondiagonal magnetic moments are of a similar magnitude.
Both of these cases seem unnatural.


{\bf MSW Mixing:\ }
The above discussion must be modified for matter-enhanced oscillations
(the MSW effect).  
First, the initial composition of the beam 
is governed not by the vacuum mixing angle $\theta_v$, but
by the initial matter mixing angle $\theta_m$, which depends on
$\Delta m^2/2E_\nu$ and the electron density.  If the initial density
is well above the resonance density, as is true for the standard
solutions to the solar neutrino problem, then $\theta_m \simeq \pi/2$
to an excellent approximation.  Then initially, $|\nu_e\rangle = \cos
\theta_m |\nu_1\rangle + \sin\theta_m |\nu_2\rangle \simeq
|\nu_2\rangle$.

Second, although a nearly pure $|\nu_2\rangle$ is produced in the
solar center, if the passage through the resonance is nonadiabatic,
then the final beam can be a mixture of $|\nu_1\rangle$ and
$|\nu_2\rangle$.  Most generally~\cite{Beacom},
the mass eigenstates evolve as
\begin{eqnarray}
|\nu_1\rangle & \rightarrow &
c_1 e^{+i\phi_a} |\nu_1\rangle + c_2 e^{+i\phi_b} |\nu_2\rangle \\
|\nu_2\rangle & \rightarrow & 
- c_2^* e^{-i\phi_b} |\nu_1\rangle + c_1^* e^{-i\phi_a} |\nu_2\rangle\,,
\label{eq:MSW}
\end{eqnarray}
where $|c_1|^2 + |c_2|^2 = 1$.  The phases $\phi_a$ and $\phi_b$ (real
functions that depend on integrals of the instantaneous mass basis
eigenvalues) are irrelevant here, due to the non-interference of different
mass eigenstates in the magnetic scattering.  For the adiabatic case
(e.g., the solar large-angle solution~\cite{Bahcall}), $c_2 = 0$.  For
the nonadiabatic case (e.g., the solar small-angle
solution~\cite{Bahcall}), and a narrow resonance region (which
naturally obtains), the probability of hopping from one mass
eigenstate to the other is $P_{hop} = |c_2|^2$, which depends on the
neutrino energy but not the distance from the source, e.g., for an
exponential density with density scale height $r_s$,
\begin{equation}
P_{hop} = 
\exp\left[-\pi\frac{\Delta m^2}{2 E_\nu} r_s (1 - \cos{2\theta_v})\right]\,.
\end{equation}

Thus for two-flavor Dirac mixing with only diagonal moments, we obtain
for the effective magnetic moment
\begin{eqnarray}
\mu_e^2 & = & |c_2|^2 |\mu_1|^2 + |c_1|^2 |\mu_2|^2 \nonumber \\
& = & P_{hop} |\mu_1|^2 + (1 - P_{hop}) |\mu_2|^2\,.
\label{eq:DirMSW}
\end{eqnarray}
Note that this is different from Eq.~(\ref{eq:Dir}), even in the
adiabatic case.  However, for the two-flavor Majorana case, we again
obtain $\mu_e^2 = |\mu_{12}|^2$, as in
Eq.~(\ref{eq:Maj}).  In both cases, since the initial state is
a pure $|\nu_2\rangle$, there are no interference terms that depend on
distance.


{\bf SK Data:\ }
The best direct limit on the neutrino magnetic moment, $1.8 \times
10^{-10} \mu_B$ at 90\% CL \cite{Derbin}, comes from studies of
neutrino-electron scattering with reactor antineutrinos.  (See
Ref.~\cite{Raffelt} and references therein for the astrophysical
limits).  As explained above, the meaning of the measured $\mu_\nu$
using solar neutrinos and reactor antineutrinos could in principle be
different.  Nevertheless, it is important to realize that
a magnetic moment numerically equal to the current reactor
limit would have a statistically significant effect on the solar
neutrino data from SK.  Since there is,
as explained below, no evidence in the data
for a nonvanishing magnetic moment, we derive, with a minimum
of assumptions,  a limit on what we call $\mu_e^{sol}$.

If the expected weak scattering rate were known (as assumed in
Ref.~\cite{Pulido}), an observed excess in the total rate would
indicate a nonzero magnetic moment.  However, as the total weak rate
is not known a priori, we instead look at the {\it shape} of the
electron spectrum for the effects of a magnetic moment.
The signature of a nonvanishing magnetic moment would be 
an enhancement, compared to the weak scattering alone, 
of the events at low recoil energies, with less enhancement
at higher energies. That is not observed.
Instead, as shown below,
the electron spectra recorded by SK have, within the
statistics, the shape that one expects from weak scattering alone (we
show below that the deviations observed currently in the highest
energy bins are irrelevant for our purpose).  However, the total
number of events is less than the standard solar model predicts,
presumably due to neutrino oscillations.  We do not need to know the
value, or the mixing mechanism behind it, of this overall reduction of
the scattering rate.

We assume only that the shape of the measured spectrum is not due to a
fortuitous cancellation between a magnetic moment effect rising at low
energies and an oscillation effect rising at high energies.  The
Sudbury Neutrino Observatory 
will check the spectral shape and total flux of the
$\nu_e$ component.

The procedure we adopt uses the measured $relative$ errors by SK and
the fact that the measured shape agrees with expectations.  We
calculate $\langle d\sigma/dT \rangle$ by folding Eq.~(\ref{eq:dsig})
with the $^8{\rm B}$ neutrino spectrum from Ref.~\cite{Boron8}.  For
both weak and magnetic scattering, we include the SK energy
resolution~\cite{SKres}, though it makes little difference in the
final results.  We histogram the results in 0.5 MeV bins in total
electron energy, as in SK.  Thus, as a function of the bin number $i$,
we have constructed the expected spectra $n_W(i)$ and $n_M(i)$ for
weak and magnetic scattering, respectively.

\begin{figure}[t]
\epsfxsize=3.25in \epsfbox{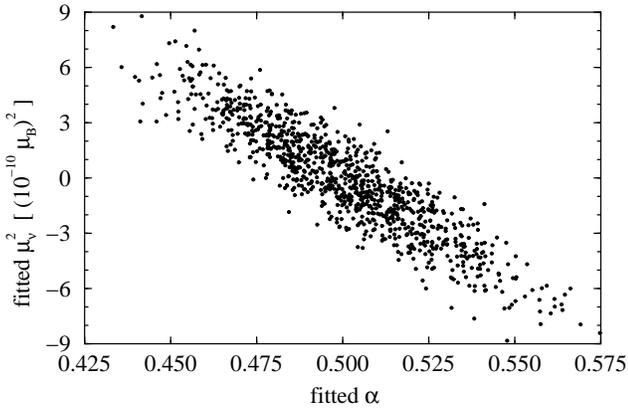}
\caption{The scatter plot illustrating the ranges and frequencies with
which the fitted $\alpha$ and $\mu_{\nu}^2$ appear. The plot
corresponds to $\alpha_{ref}=0.5$ and $\mu_{ref}=0$, and uses the
relative errors (504-days data) of Ref.~\protect\cite{Fukuda}.}
\label{fig:scatt}
\end{figure}

\begin{figure}[t]
\epsfxsize=3.25in \epsfbox{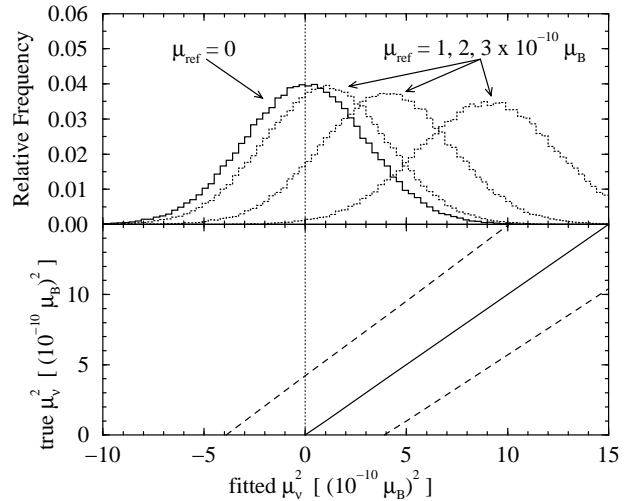}
\caption{The upper panel shows the relative frequencies of different
fitted $\mu_{\nu}^2$ values for the indicated values of $\mu_{ref}$
and for $\alpha_{ref}=0.5$.  The calculation uses the relative errors
(504-days data) of Ref.~\protect\cite{Fukuda}. 
The lower panel allows one to find for any
fitted $\mu_{\nu}^2$ the most probable value of the true $\mu_{\nu}^2$
(full line) and the 10\% and 90\% CL interval of that quantity (dashed
lines).}
\label{fig:freq}
\end{figure}

In order to determine the upper limit of $|\mu_e^{sol}|^2 $ we must
take into account the statistical fluctuations in the SK data.  While
the data points divided by the solar model expectation are consistent
with an energy independent reduction factor $\alpha$, the individual
bins are distributed, presumably randomly, around that value.  To take
that into account we choose some reference values $\alpha_{ref}$ and
$\mu_{ref}$ and create a set of $simulated$ data, $n_S(i)$, which are
Gaussian-distributed around the theoretical expectation
$\alpha_{ref}n_W(i) + \mu_{ref}^2n_M(i)$ with the relative errors
$\sigma(i)$ given by SK.

We then minimize the $\chi^2$,
\begin{equation}
\chi^2 = \sum_i \left[ \frac {\alpha n_W(i) + \mu_{\nu}^2 n_M(i) - n_S(i)}
{\sigma(i) n_S(i)} \right]^2\,.
\label{eq:chi2}
\end{equation}
with respect to the fit parameters $\alpha$ and $\mu_\nu^2$.  For
fixed $\alpha_{ref}$ and $\mu^2_{ref}$, we repeat this procedure many
times and plot the frequencies with which given values of the fitted
$\alpha$ and $\mu_\nu^2$ appear.  An example of the scatter plot of
the fit parameters is shown in Fig.\ref{fig:scatt}.  One can see,
naturally, that the most probable values of the fit are the reference
values $\alpha_{ref}$ and $\mu_{ref}$.  Also, the two variables are
strongly anticorrelated (correlation
coefficient $r \simeq -0.9$), i.e., larger $\alpha$ is accompanied by
smaller $\mu_{\nu}^2$.
Dividing the numerator and denominator in Eq.~(\ref{eq:chi2}) by
$n_W(i)$, one sees that the $\chi^2$ depends only on $r_i = n_S(i)/n_W(i)$,
i.e., precisely on the quantities published by SK \cite{Fukuda}.

By repeating the calculation for different values of $\mu_{ref}$ and
projecting on the $\mu_{\nu}^2$ axis, one gets the distributions shown
in the upper panel of Fig.~\ref{fig:freq}.  These distributions are
Gaussian, and their width is almost independent of the chosen value of
$\mu_{ref}$.  Based on them we obtain the lines in the lower panel of
Fig.~\ref{fig:freq} signifying confidence levels at 10\%, 50\% (the
mean), and 90\%.  For a given fitted $\mu_\nu^2$ obtained in an
experiment, these allow one to determine the likely range of the true
$\mu_{\nu}^2$.  For example, if one found a fitted $\mu_{\nu}^2 \simeq 4$, 
then from Fig.~\ref{fig:freq}, the most probable true value is
$\mu_{\nu}^2 \simeq 4$, with the upper limit being $\simeq 8$ and
the lower limit being $\simeq 0$.  
Similarly, for a fitted $\mu_\nu^2 \simeq 0$, 
the true  $\mu_{\nu}^2$ is $ \lesssim 3.9$.  
This would be the
largest true $\mu_{\nu}^2$, due to statistical fluctuations of
the finite data, that could have given this fitted $\mu_{\nu}^2 \simeq 0$.

In this way we solve problems associated with the statistical
fluctuations as well as with the constraint $|\mu_e^{sol}|^2 \ge 0$.
Figure~2 was calculated with $\alpha_{ref} = 0.5$, as observed in SK, but
doesn't change significantly for $0.4 < \alpha_{ref} < 0.6$.  Note that the
results summarized in Fig.~\ref{fig:freq} can be also obtained
analytically, without generating many simulated spectra. The
conclusions, in particular the lower panel of Fig.~\ref{fig:freq},
simply follow from the properties of the individual sums in
Eq.~(\ref{eq:chi2}).

Using the SK data~\cite{Fukuda}, the fitted $\alpha \simeq 0.5$; the
exact value is irrelevant since we are testing only the spectral
shape, and not the normalization.  The fitted values of $\mu_\nu^2$
are slightly (but not significantly; see Fig.~2) negative: $\simeq
-5$, $-3$, and $-2$ (in the units of Table~I) for the 504-, 708-, and
825-days data sets.  The slight (but diminishing with time) positive
slope observed in the data cannot be caused by a magnetic moment
(which causes an increase at low energies), though it could be caused
by oscillations.  The most conservative conclusion is therefore to say
that the slope is {\it not negative}, i.e., that the fitted
$\mu_\nu^2$ values are {\it not positive}.  That is, we obtain
the limit by using Fig.~2 (and its analogs) and an assumed fitted
value of $\mu_\nu^2 = 0$.  Thus the limits in Table~I
are slightly weaker than what is naively implied by the data, but are
more robust.  The sensitivity to $ |\mu_e^{sol}|$ improves with time
only as $t^{-1/4}$, but the addition of more low-energy bins (e.g., the
two added since the 504-days data) gives a more dramatic improvement.

The uncertainties $\delta\alpha$ in Table I reflect the increase in
the error in the parameter $\alpha$ when one allows magnetic
scattering.  Our procedure does not include the correlations between
systematic errors in different bins and therefore will not reflect the
full systematic uncertainty.  Note that in the standard
analysis~\cite{Fukuda} one assumes $\mu_{\nu} = 0$ and hence the
uncertainty $\delta\alpha$ is reduced by the factor $1/\sqrt{1 - r^2}
\simeq 2.3$.  Similarly, if the value of $\alpha$ were accurately and
independently known, and we fit for $\mu_{\nu}^2$ only, an identical
improvement in the upper limit of $|\mu_e^{sol}|^2$ would result.

\begin{table} 
\caption{Limits on the magnetic moment (at 90\% CL, in units of
$10^{-10} \mu_B$) and statistical errors (1$\sigma$) on $\alpha$ for
various data sets~\protect\cite{Fukuda}: 504 days (I); the
same just up to 13.0 MeV, to exclude the bins with an apparent
excess compared to the constant data/SSM ratio (II); 708 days; and 
825 days.  In all cases we ignore the bin from 14.0 -- 20.0 MeV.}
\begin{tabular}{l|r|r|r}
case		& $|\mu_e^{sol}|^2$ & $|\mu_e^{sol}|$ & $\delta\alpha$ \\
\hline
504 days, I	& $\le 3.9$	& $\le 2.0$	& 0.025 \\
504 days, II	& $\le 4.2$	& $\le 2.0$	& 0.027 \\
708 days 	& $\le 2.5$	& $\le 1.6$	& 0.018 \\
825 days	& $\le 2.3$	& $\le 1.5$	& 0.017 \\ 
\end{tabular}
\end{table}


{\bf Conclusions:\ }
In this paper, we present three new results.
{\it First}, that while neutrino magnetic moments are most fundamentally
defined for mass eigenstates, in several cases of practical interest
non-oscillating (i.e., independent of distance) effective magnetic
moments can be defined for the flavor eigenstates.  For Dirac
neutrinos with only diagonal moments, these results are
Eq.~(\ref{eq:Dir}) for vacuum mixing and Eq.~(\ref{eq:DirMSW}) for MSW
mixing.  For Majorana neutrinos with two flavors, the result is
$\mu_e^2 = |\mu_{12}|^2$, for either vacuum or MSW mixing.
{\it Second}, that MSW mixing can change the definition of the effective
magnetic moment (allowing a dependence on the neutrino energy), so
that the measured moments using $\bar{\nu}_e$ from a reactor and
$\nu_e$ from the Sun could be different.
{\it Third}, that the {\it shape} of the SK recoil electron spectrum can be
used to place a limit on the neutrino magnetic moment (note that we do
not invoke any mechanism for neutrino interaction with the solar
magnetic field).  In general, this is a new limit, independent of the
limit from reactor studies (with the same meaning only for solar
vacuum oscillations).  In any case, the limit obtained using the
preliminary 825-days data, $|\mu_e^{sol}| \le 1.5 \times 10^{-10}
\mu_B$, is comparable to the existing reactor
limit of $1.8 \times 10^{-10} \mu_B$ \cite{Derbin}.


This work was supported in part by the U.S. Department of Energy under
Grant No. DE-FG03-88ER-40397.  J.F.B. was supported by Caltech.  We
thank Boris Kayser, Bob Svoboda, and Mark Vagins for discussions, and
the SuperKamiokande collaboration for supplying the 825-days results.



\end{document}